\documentclass[runningheads,a4paper]{llncs}
\usepackage{times,amsmath,epsfig}

\usepackage{indentfirst}

\usepackage{algpseudocode}
\usepackage{algorithm}
\usepackage{mathrsfs}

\usepackage{paralist}

\def\actors{{\mathbf A}}
\def\topics{{\mathbf O}}

\title{\emph{MAP:} Microblogging Assisted Profiling of TV Shows}
\author{%
{Xiahong Lin{\small $~^{1}$}, Zhi Wang{\small $~^{2}$}, Lifeng Sun{\small $~^{1}$} }%
\institute{$^{1}$\,Tsinghua National Laboratory for Information Science and Technology\\
Department of Computer Science and Technology, Tsinghua University\\
$^{2}$\,Graduate School at Shenzhen, Tsinghua University\\
\{lin-xh12@mails., wangzhi@sz., sunlf@\}tsinghua.edu.cn
}
}

\begin{document}
\maketitle

\begin{abstract}

Online microblogging services that have been increasingly used by people to share and exchange information, have emerged as a promising way to profiling multimedia contents, in a sense to provide users a socialized abstraction and understanding of these contents. In this paper, we propose a microblogging profiling framework, to provide a social demonstration of TV shows. Challenges for this study lie in two folds: First, TV shows are generally \emph{offline}, i.e., most of them are not originally from the Internet, and we need to create a \emph{connection} between these TV shows with online microblogging services; Second, contents in a microblogging service are extremely noisy for video profiling, and we need to strategically retrieve the most related information for the TV show profiling. To address these challenges, we propose a \emph{MAP}, a microblogging-assisted profiling framework, with contributions as follows: i) We propose a joint user and content retrieval scheme, which uses information about both actors and topics of a TV show to retrieve related microblogs; ii) We propose a social-aware profiling strategy, which profiles a video according to not only its content, but also the social relationship of its microblogging users and its propagation in the social network; iii) We present some interesting analysis, based on our framework to profile real-world TV shows.



\end{abstract}


\section{Introduction}

Recent years have witnessed an increasing popularity of online microblogging services (e.g., Twitter and Weibo), and the rapid convergence between the online microblogging service and other multimedia services including online video streaming \cite{xu2010cuckoo}. This new trend makes the online microblogging service a promising way to \emph{profiling} multimedia contents, i.e., using the information from the microblogging systems to abstract, demonstrate and enrich multimedia contents \cite{holotescu2011m3}. In this paper, we study profiling one of the most important multimedia types --- TV shows (videos) that are generally received by users on TVs, with online microblogging services.


Profiling TV shows is important to both content providers and users \cite{wakamiya2011twitter}. On one hand, an interesting profile of a TV show will allow a content provider to demonstrate the TV show more effectively, e.g., more users can be attracted to watch the TV show, indicating more subscriptions from the content provider. On the other hand, an illustrating profile of a TV show also gives a user an efficient way to know about the TV show before actually watching it, improving the whole show viewing experience.

Today, it has become a norm rather than an exception, for users to share and receive information and use online social networks to search multimedia contents \cite{benevenuto2009characterizing}. In this personal and social context, even TV show profiles have to be highly personalized and social-aware. Traditional video profiling is generally based on video content only \cite{dimitrova2002applications}, e.g., the profile is composed by the video title, actor information, etc., and has the following fundamental drawbacks: (1) For a content provider, the video profile is generated without any consideration about the \emph{social effect} the profile may make, e.g., how the video is going to attract users in the online social network and how the profile will be correlated with other videos. (2) For users, the profile is \emph{static} and composed in an one-size-fit-all manner, i.e., the same video profile is provided to all users without any user-specific information.

To address these challenges, we propose a TV show profiling framework called MAP by exploring the new design space of using online social network information. Our contributions in this paper are summarized as follows: (1) We design a joint user and content information retrieval scheme. We use information about both actors and topics in a TV show to retrieve microblogs from a microblogging system. (2) We propose a social-aware profiling scheme, which profiles a TV show according to not only its content, but also the social relationship of its microblogging users and its propagation in the social network. (3) We present some interesting results and analysis, using our framework to profile real-world TV shows.

The rest of this paper is organized as follows. We present the information retrieval strategy and the microblogging-assisted profiling framework in Sec.~\ref{sec:profile}. Based on this framework, we present several analytical results in Sec.~\ref{sec:measurement}, using data from real-world TV show and microblogging systems. Finally, we conclude the paper in Sec.~\ref{sec:conclusion}.


\section{Profiling TV Shows with Microblogs}
\label{sec:profile}

\begin{table*}
    \begin{center}
    \caption{Representative TV shows studied.}
	\label{tab:videolist}
    \begin{tabular}{|c|c|c|c|r|r|}
    \hline
    \textbf{ID} & \textbf{TV Show title} & \textbf{Category}  & \textbf{Leading Actors} & \textbf{\# microblogs} & \textbf{\# microblog users}\\
	\hline
	  v1 & \textit{Aiqing Shuixingle} & Love/Idol &  Stephy Qi/Tiffany Tang & $571,227$ & $388,109$ \\
	  v2 & \textit{Bubu Jingxin} & Love/Costume & Shishi Liu/Nicky Wu & $2,748,223$ & $1,772,973$ \\
	  v3 & \textit{Meiren Xinji} & Love/Costume & Ruby Lin/Mi Yang & $169,513$ & $147,812$ \\
	  v4 & \textit{Nanrenbang} & Idol/Modern & Di Yao/Lei Huang & $1,141,362$ & $844,655$ \\
	  v5 & \textit{Qianshan Muxue} & Love/Idol &  Yinger/Hawick Lau & $271,371$ & $190,288$ \\
	  v6 & \textit{Qingshi Huangfei} & Love/Costume &  Ruby Lin & $729,010$ & $544,439$ \\
	  v7 & \textit{Xuebao} & War/Love & Zhang Wen & $83,835$ & $67,561$ \\
	  v8 & \textit{Wuxiekeji} & Love/Idol & Stephy Qi/Tiffany Tang & $158,534$ & $116,436$ \\
      v9 & \textit{Aiqing Gongyu} & Comedy/Love & Loura/Jean Lee & $1,036,144$ & $775,171$ \\
	  v10 & \textit{Aishang Zhameile} & Love/Idol &  Cyndi Wong & $78,522$ & $59,080$ \\
	  v11 & \textit{Shachun} & Family/Tragedy &  Helen Tao & $56,524$ & $47,582$ \\
	  v12 & \textit{Duanci} & War/Suspense &  Lei Tong & $14,509$ & $11,410$ \\
	  v13 & \textit{Duanhounu} & War/Historical &  Yunxiang Gao & $5,370$ & $4,260$ \\
	  v14 & \textit{Huangtuteng} & Love/Costume & Hans Zhang/Bing Bai & $66,881$ & $48,400$ \\
	  v15 & \textit{Luohun Shidai} & Love/Family &  Zhang Wen/Di Yao & $3,544,006$ & $2,789,363$ \\
	 \hline
\end{tabular}
\end{center}
\end{table*}

In this section, we present our profiling framework.
First, to effectively utilize the noisy microblog information, we retrieve the most informative microblogs for a TV show, by jointly considering actors and topics of the show; Second, to efficiently provide a profile both informative to a TV show content provider and users, we create a profile for TV shows, from the perspective of user, content, social relationship and propagation.


\subsection{Background and Data Collection}

We have collected data from a TV show service provider \textit{BesTV} and an online microblogging service provider \textit{Tencent Weibo} for our study. For simplicity, we refer to the TV show service as V, and the microblogging service as M. We have collected TV shows from the TV show company V. Our dataset contains $233,804,070$ valid user viewing records of $69$ TV shows during June 1 through December 30, 2011. We have also collected the meta information of these TV shows, including the title, category, and the actors in the TV shows. In Table \ref{tab:videolist}, we list the representative TV shows used in our study. We will use the indices (v1, v2, \ldots, v15) to denote these TV shows in this paper.


\subsection{Informative Microblogs Retrieval}
\label{sec:retrieval}

Microblogs are often intrinsically noisy, as users have been using such online microblogging services to exchange all kinds of information. Besides, among these microblogs posted by users, only a little fraction of them are about TV shows, e.g., in popular microblogging systems including Twitter, only a very small fraction of microblogs are about TV shows\footnote{http://en.wikipedia.org/wiki/Twitter}. It is a must to find the most informative contents when we use microblogs to profile TV shows. We propose to jointly use both actors and topics of a TV show to retrieve the most informative microblogs.

%
%

\subsubsection{Microblogs from Actors.}

In our study, we focus on the professional TV shows, which are produced by large companies, instead of user-generated videos \cite{milliken2008user}, as such professional videos (e.g., a movie or TV show) are the mainstream type of videos on a video service. For such a TV show $v$, we are able to find an actor list $\actors_v$, in which actors of the show are included. According to our measurement results, there is over $1/3$ of the leading actors having a microblogging account on average. We collect microblogs posted by accounts in $\actors_v$, if they have social accounts. For example, a TV show $v6$ in our study named ``\textit{Qingshi Huangfei}'' has an actor named ``Ruby Lin'', who has a social account ID on the microblogging system M called ``linxinru'', and we have collected microblogs posted by this ID for profiling this TV show.


\subsubsection{Microblogs from Topics.}

For a TV show $v$, we are also able to find a set of topics $\topics_v$, including the title of the TV show, category, abstract, etc. In our study, such topics are used to search for microblogs, in a keyword-matching manner, i.e., if the content of a microblog contains any keyword in $\topics_v$, this microblog will be used for profiling TV show $v$. For example, the same TV show $v6$ ``\textit{Qingshi Huangfei}'' has the leading roles ``Mafuya'', ``Mengqiyou'', ``Liuliancheng'', which are used as the topics to retrieve microblogs.




Using the above microblogs as ``seed'' microblogs, we further collected microblogs that are repost and comments to them. In summary, we have collected overall $23,963,680$ valid microblogs posted by $11,685,768$ users in M. In Table \ref{tab:videolist}, we show the number of microblog users and microblogs for each of the representative TV shows. Each microblog record contains the follows ID, name, IP address of the publisher, time stamp when the microblog was posted, IDs of the root microblog if it is a reposted microblog, and contents of the microblog. Interested readers are referred to our previous studies on collecting the microblog traces \cite{zhi-acmmm2012,zhi-tmm2013}.

\subsection{MAP: Microblogging-Assisted Profiling of TV Shows}

Based on microblogs, user profiles, and social connections collected using the retrieval scheme above, we profile TV shows from the following aspects. (1) \emph{User aspect:} We present the demographics of microblogging users of a TV show, and their preferences; (2) \emph{Content aspect:} We present the sentiment analyses of TV shows by the microblog content posted to the shows, and the ``social network'' of TV shows; (3) \emph{Social relationship aspect:} We present the social topology of microblogs views of a TV show, as well as the social network and social influence of actors of TV shows, who have online microblogging accounts; (4) We present the propagation of microblogs users across different TV shows, i.e., how users' attention is shifting from a TV show to another.

Next, we present a case study using MAP, including the insights from the observations.

\section{Insights Learnt from MAP}
\label{sec:measurement}

In this section, we present several results based on our profiling framework, from the aspects of user, content, social relationship and social propagation.

\subsection{Demographics and Preference of Viewers}

\subsubsection{Demographics of Microblogging Viewers.}

\begin{figure}[h]
    \begin{minipage}[t]{.48\linewidth}
        \centering
            \includegraphics[width = \linewidth]{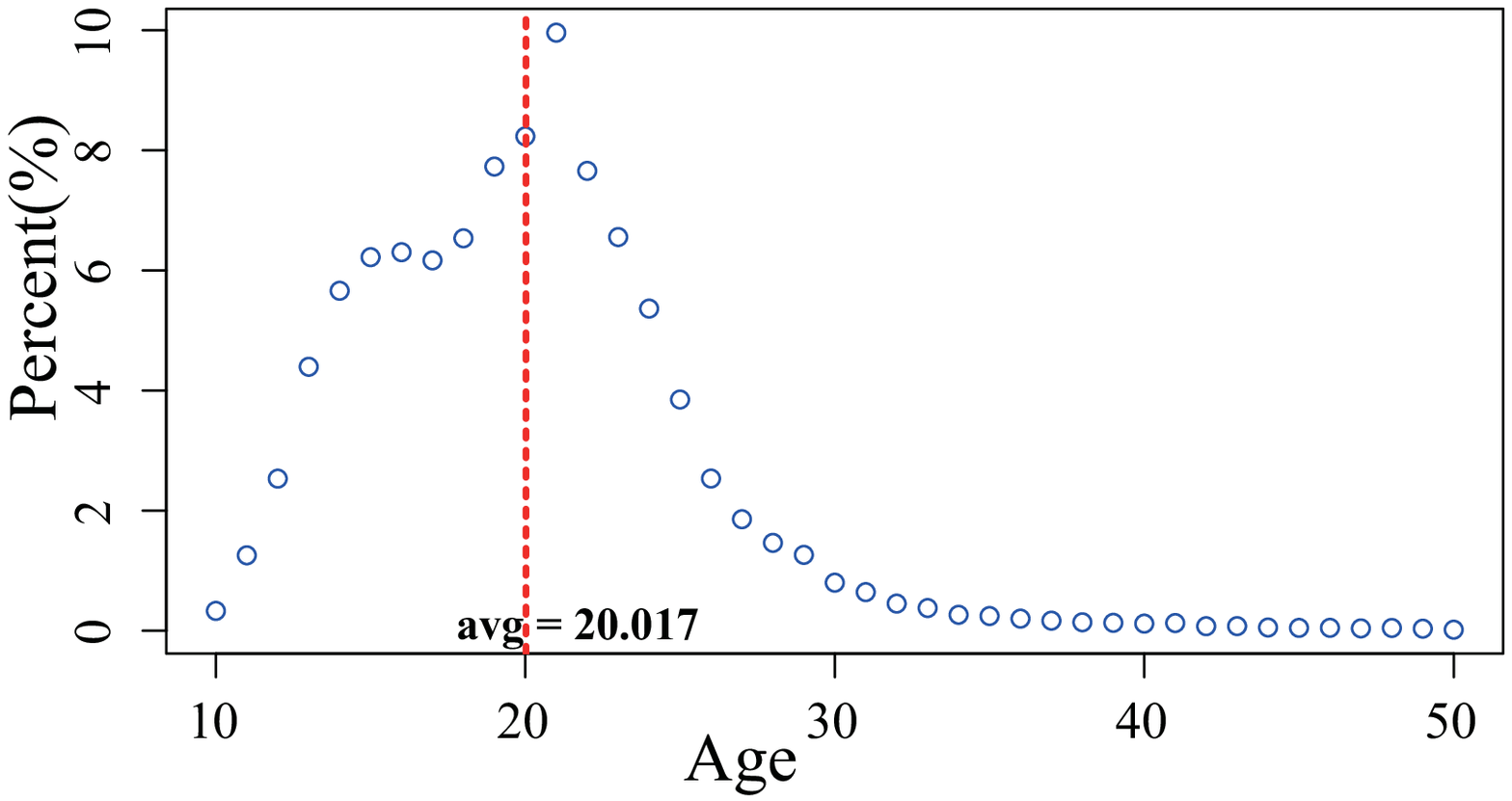}
        \centerline{\scriptsize (a) Microblogging user ages}
    \end{minipage}
    \hfill
    \begin{minipage}[t]{.48\linewidth}
        \centering
            \includegraphics[width = \linewidth]{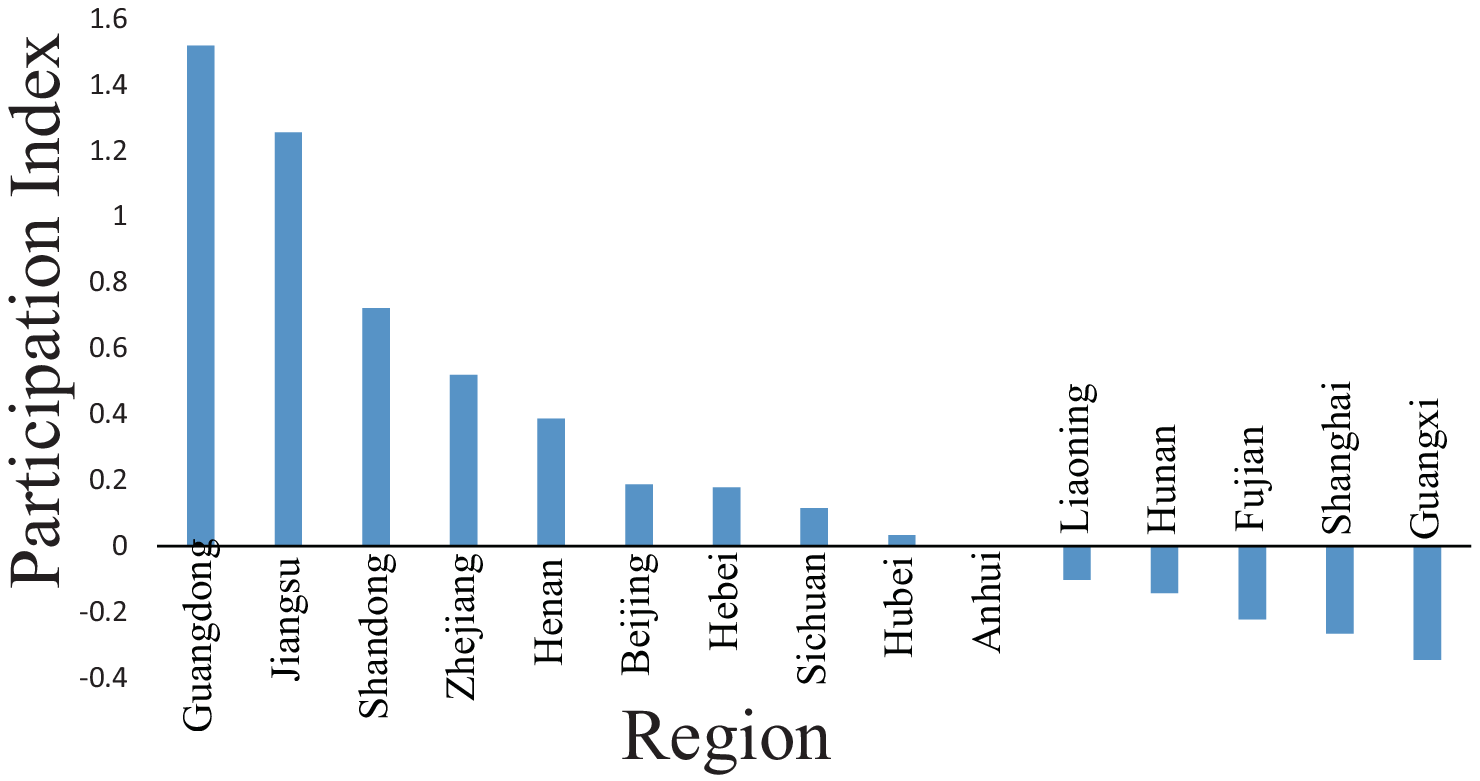}
        \centerline{\scriptsize (b) Microblogging user regions}
    \end{minipage}
    \caption{Demographic of users.}
    \label{fig:userProfile-1}
\end{figure}

Using the microblogging user information, we are able to demonstrate TV shows from a user aspect. Fig.~\ref{fig:userProfile-1}(a) shows the age distribution of normal users. It forms a normal-like distribution, and we observe that the average age is $20$, indicating the major of the users are quite young. Fig.~\ref{fig:userProfile-1}(b) shows the top $15$ province-level regions where the microblog users are located. Each bar in the figure is a top-10 participation index, defined as
\begin{equation}
    PI_i = \frac{UN_i - UN_{10}}{UN_{10}}
    \label{eqn:pii}
\end{equation}
where $UN_i$ is the numbers of users in region $i$. From this demographics profiling, we are able to show the regional distribution of users. For example, we observe that for this particular TV show provider, Guangdong Province ranks the first, i.e., it has the most microblog users. In summary, we observe that the young users in top-tier cities in China are the major viewers of these TV shows.

\subsubsection{User Preference Clustering.}

A TV show provider gives 3 labels to each TV show to categorize its content. We use such content labels to infer users' preferences of video contents. According to the microblogs posted by a user, we are able to assign each user with a weight list of content labels. Using the weighted label lists of users, we are able to calculate a ``distance'' between any two users. Then, we cluster these users using a K-means clustering algorithm \cite{hartigan1979algorithm} using based on the label distance.

\begin{figure}[h]
    \begin{minipage}[t]{.48\linewidth}
        \centering
            \includegraphics[width = \linewidth]{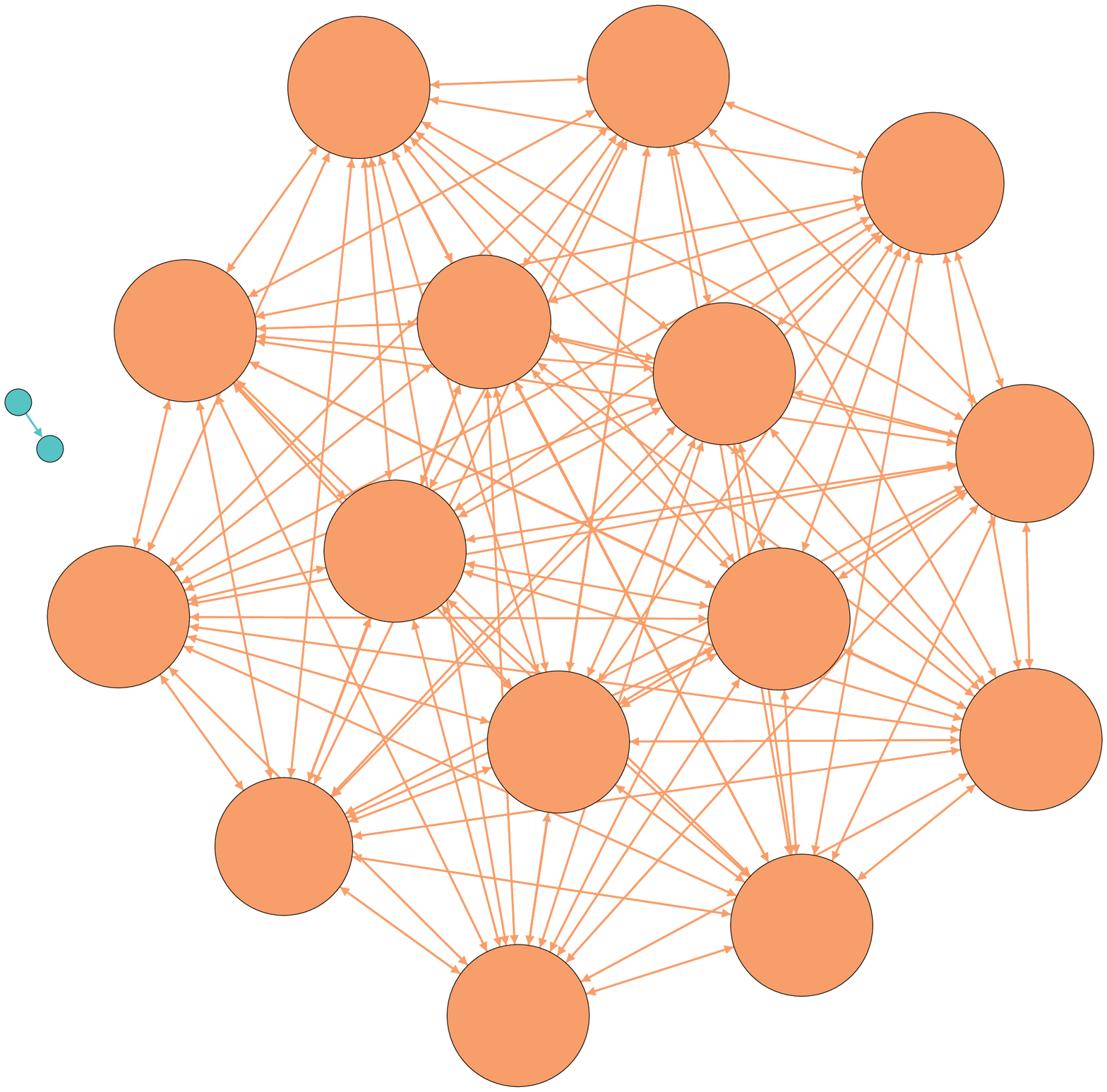}
        \centerline{\parbox[b]{\linewidth}{\scriptsize (a) Social graph with a large clustering coefficient}}
    \end{minipage}
    \hfill
    \begin{minipage}[t]{.48\linewidth}
        \centering
            \includegraphics[width = \linewidth]{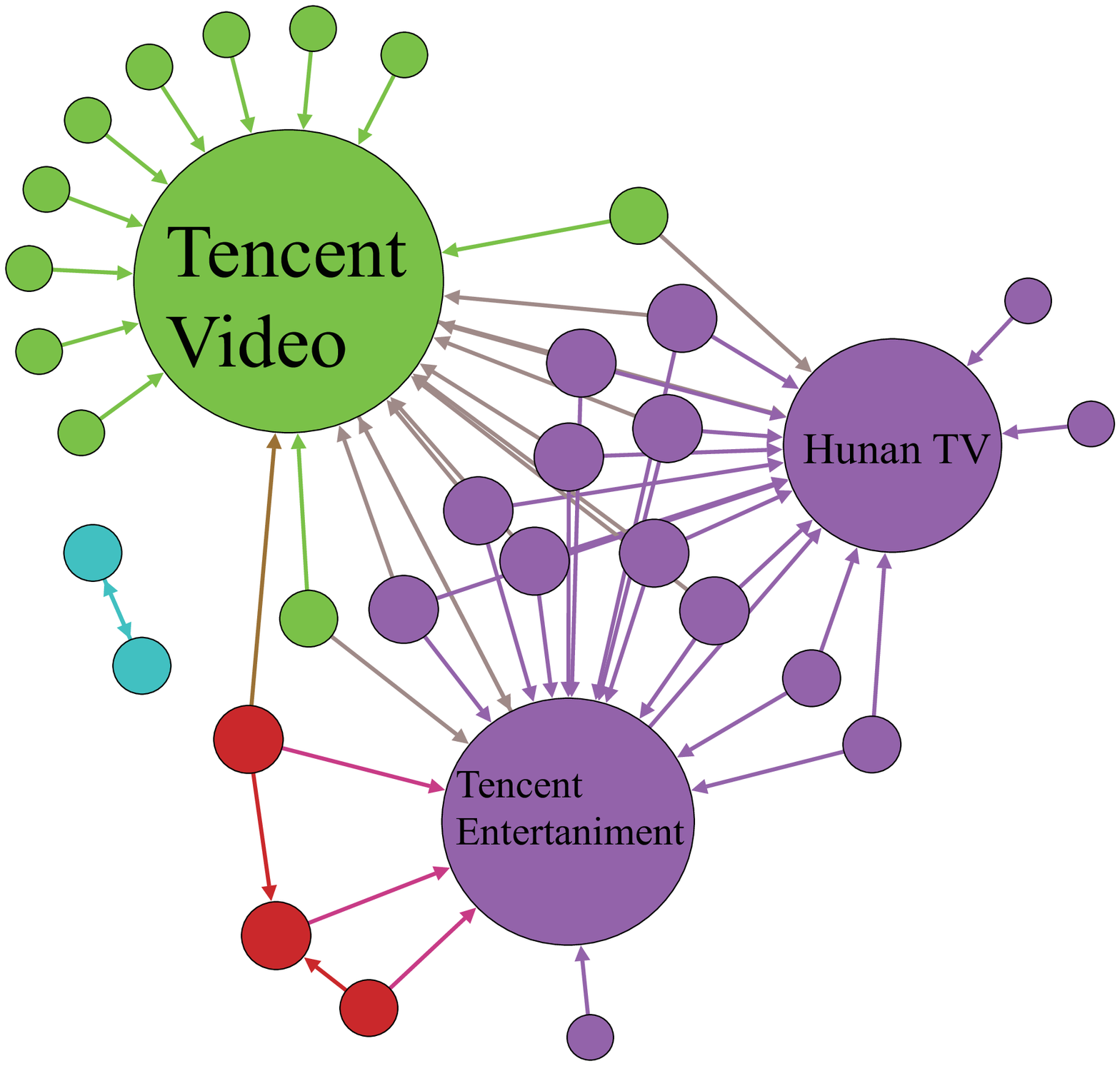}
        \centerline{\parbox[b]{\linewidth}{\scriptsize (b) Social graph with VIP users}}
    \end{minipage}
    \caption{Social relationship among users sharing similar preference.}
    \label{fig:activityCluster}
\end{figure}

In Fig.~\ref{fig:activityCluster}(a), we plot the social connections between users that clustered into one group. Each circle represents a user, and a connection between two users indicates that one of them is following the other. We observe that in this figure users are closely connected to each other with a large clustering coefficient \cite{soffer2005network}, when they are clustered into the same group by K-means according to their preference.

Using the same approach to cluster users, we further find that users' preference can be indicated from who they are socially connected in a microblogging system. In Fig.~\ref{fig:activityCluster}(b), we have plotted several VIP microblog users, i.e., \textit{Tencent Video}, \textit{Hunan TV}, and \textit{Tencent Entertainment}. We observe that when users are clustered into the same group, they only follow the VIP users in that group, but have no connection with other users in the same group. This observation indicates that a TV show provider is able to identify users that may be interested in a TV show, by looking at the followers the VIP accounts.

\subsection{Sentiment and Social Network of TV Shows}

Online microblogging services also allow us to explore the sentiment and social network of TV shows.

\subsubsection{Sentiment Analysis.}

\begin{figure}[h]
    \begin{minipage}[t]{.48\linewidth}
        \centering
            \includegraphics[width = \linewidth]{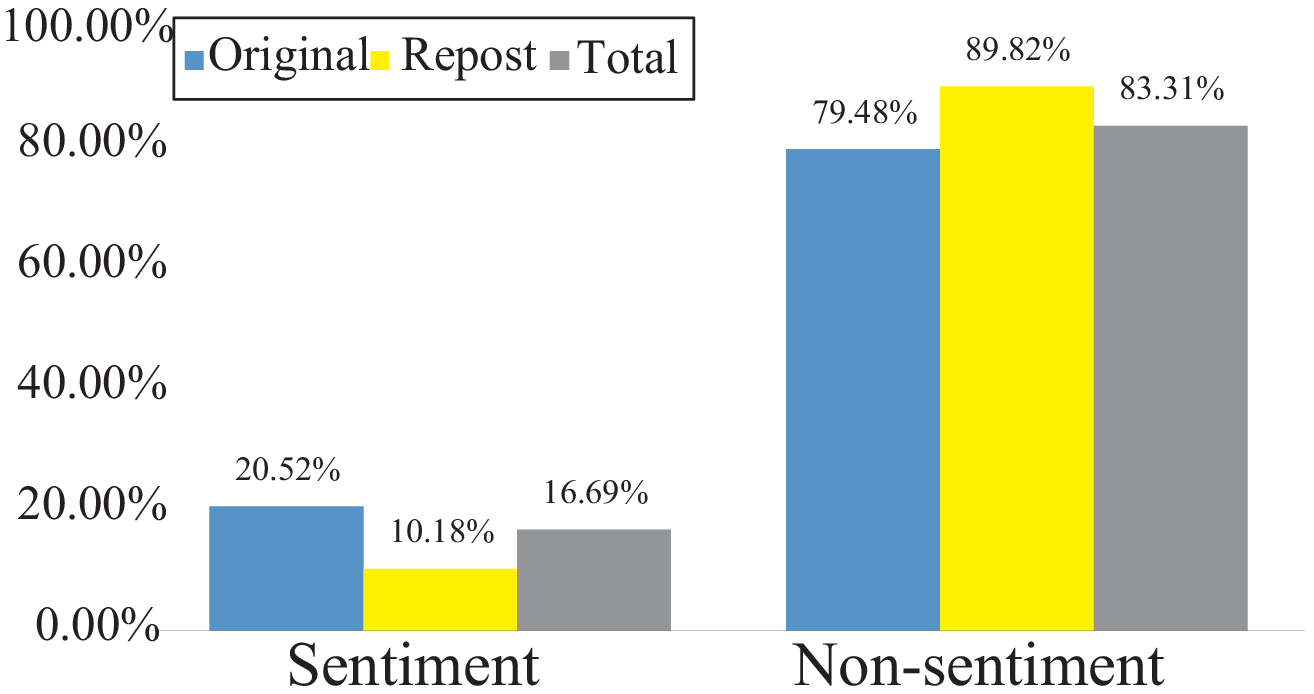}
        \centerline{\parbox[b]{\linewidth}{\scriptsize (a) Overall of posts}}
    \end{minipage}
    \hfill
    \begin{minipage}[t]{.48\linewidth}
        \centering
            \includegraphics[width = \linewidth]{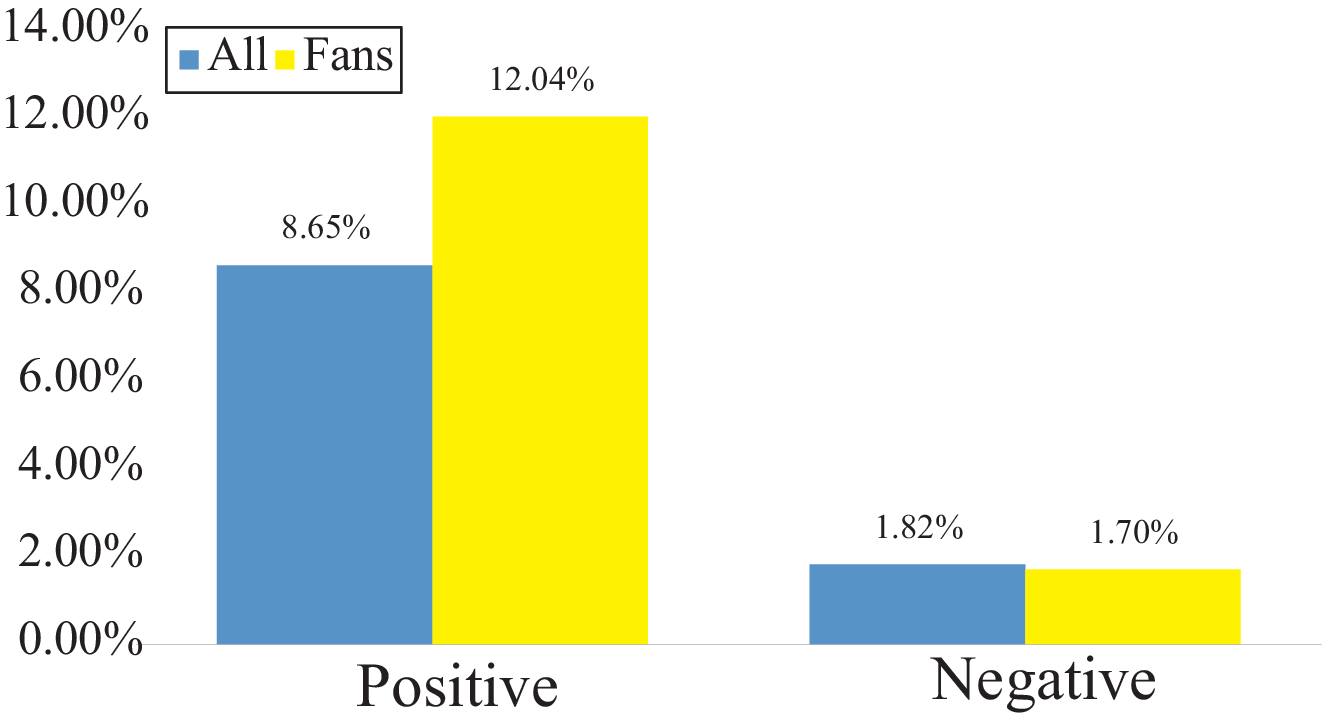}
        \centerline{\parbox[b]{\linewidth}{\scriptsize (b) Positive vs.~negative}}
    \end{minipage}

    \caption{Sentiment distribution of microblogs on TV shows.}
    \label{fig:sentiment-1}
\end{figure}

Microblogs allow us to use the content of microblogs for sentiment analysis, i.e., we label each microblog ``positive'', ``negative'', or ``non-sentiment'', according to its content. After analyzing all the posts, we get the following observations: (1) Initial microblog posts on TV shows contain more emotional contents than reposts and comments, as illustrated in Fig.~\ref{fig:sentiment-1}(a);  (2) There are much more positive microblogs on TV shows than negative ones, as illustrated in Fig.~\ref{fig:sentiment-1}(b).


\begin{figure}[h]
	\centering
		\includegraphics[width = \columnwidth]{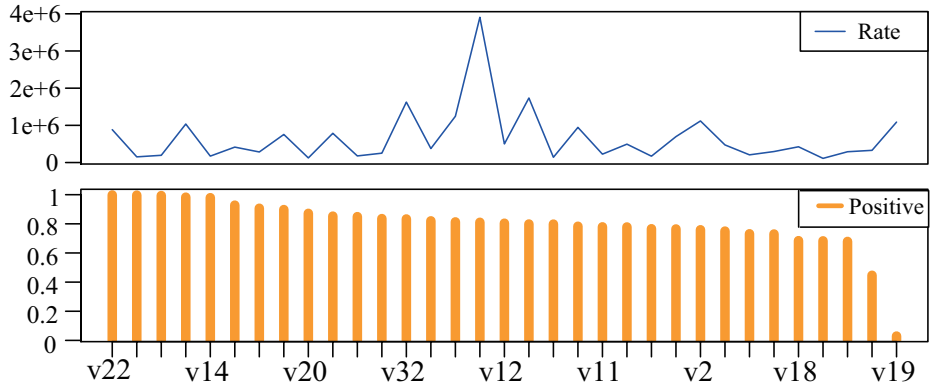}
	\caption{Sentiment distribution of different TV shows, along with their TV rating.}
	\label{fig:sentiment-2}
\end{figure}

Next, we study the sentiment distribution of microblogs in different TV shows. We find that most of them have higher positive proportions except for two TV shows. As shown in Fig.~\ref{fig:sentiment-2}, each bar is the fraction of positive microblogs versus the TV show rank, and the curve is the view number of each TV show versus the show rank. We observe that a TV show that has too many positive microblogs or too many negative microblogs tend to have a small number of views. The reason may be that highly-viewed shows will attract more diverse discussions on the microblogging system.

\subsubsection{``Social Network'' Between TV Shows.}

\begin{figure}[h]
    \begin{minipage}[t]{.48\linewidth}
        \centering
            \includegraphics[width = \linewidth]{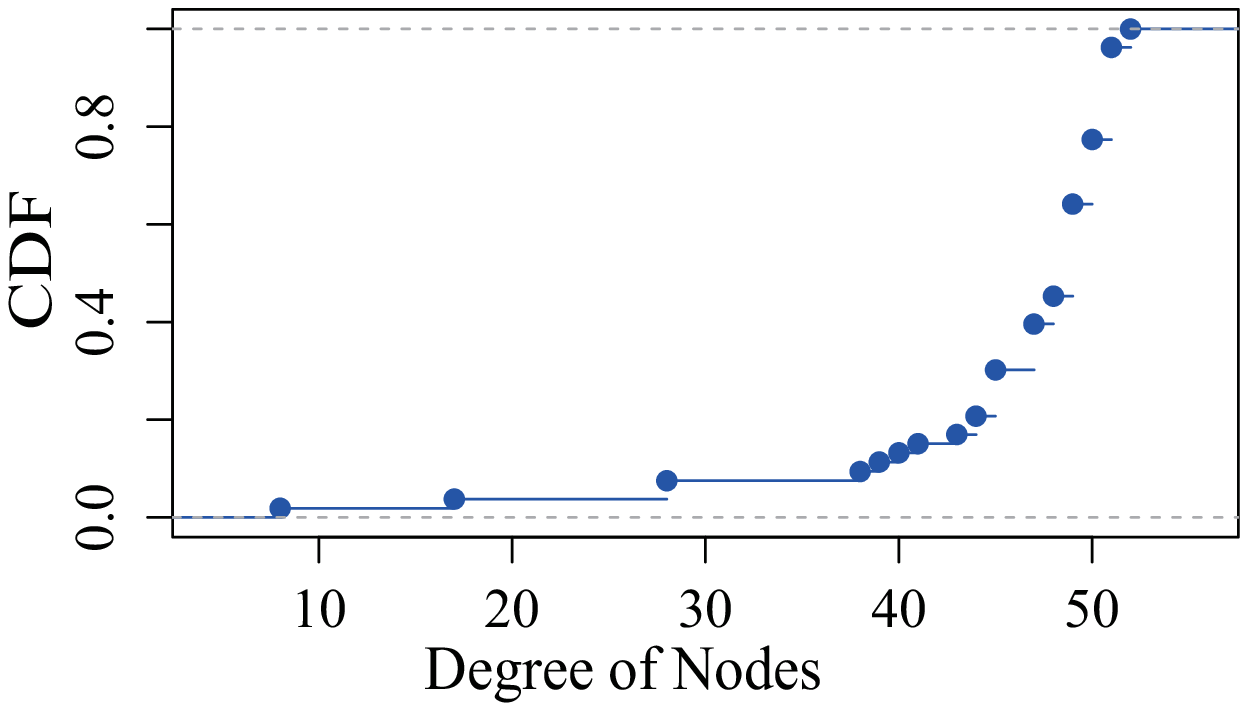}
        \centerline{\scriptsize (a) Degree distribution}
    \end{minipage}
    \hfill
    \begin{minipage}[t]{.48\linewidth}
        \centering
            \includegraphics[width = \linewidth]{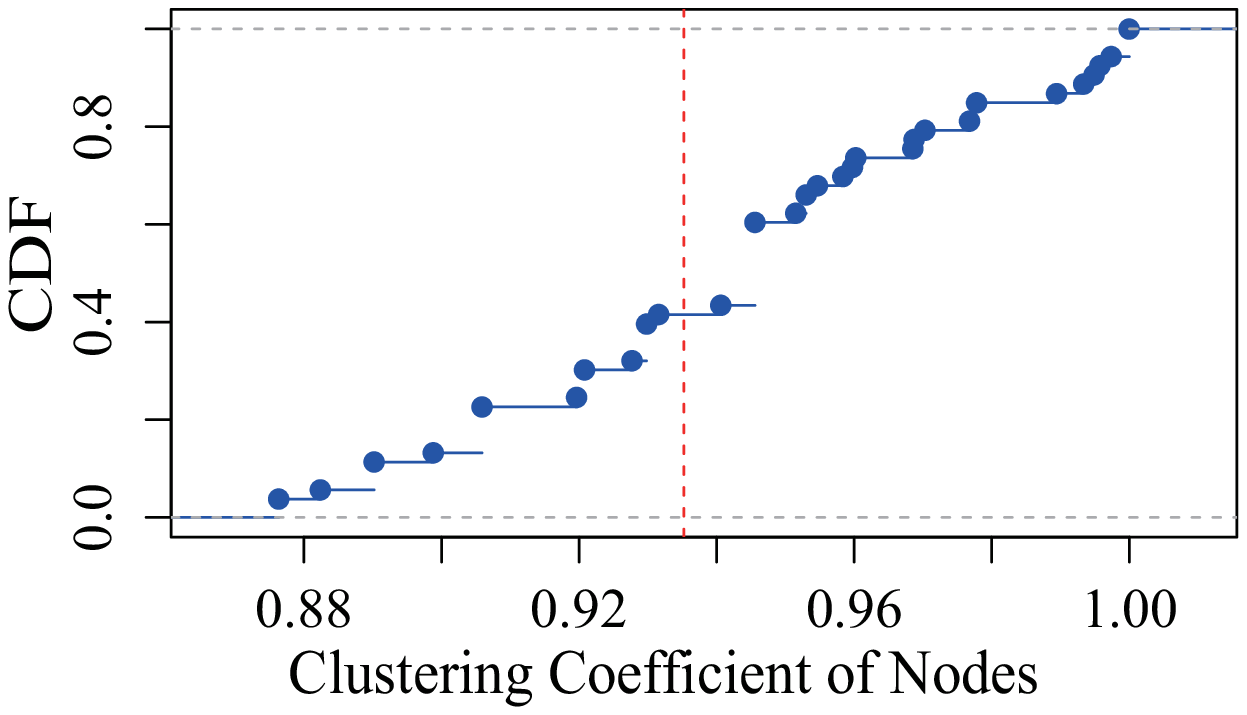}
        \centerline{\scriptsize (b) Local clustering coefficient distribution}
    \end{minipage}
    \caption{Statistics of Nodes.}
    \label{fig:degreeAndClustering-p}
\end{figure}

Users can post microblogs which relate to more than one TV shows, therefore, TV shows may be connected via their common microblog users. Using this idea, we create a social network of TV shows, i.e., two shows are connected if they share a microblog user. Fig.~\ref{fig:degreeAndClustering-p} shows the statistics of nodes (TV shows). Fig.~\ref{fig:degreeAndClustering-p}(a) illustrates the degree distribution of TV shows. We observe that about $90\%$ of the shows have a degree higher than $40$, and the mean is $45.8$, indicating that the graph is almost fully-connected. Fig.~\ref{fig:degreeAndClustering-p}(b) illustrates the CDF of local clustering coefficients of the shows. We note that TV show graphs with higher level clustering coefficient indicate more common users.

Next, we study the community structure of the social network of TV shows, as illustrated in Fig.~\ref{fig:community-p}. Since the average clustering coefficient of the network is $0.94$, which is 2x of the same scale of a random network. The average diameter of the graph is $1.12$, which is at the same scale of a random network. Also, we study its modularity using a heuristic approach \cite{blondel2008fast}. We have modularity value of $0.05$ (Using the community structure detecting algorithm proposed by Du et al.~\cite{duhaifeng}) for the same social network of the TV shows. The very small value indicates that the social network of TV shows does not have a clear community structure \cite{newman2003structure}.

Furthermore, we explore the degree distribution of the network. Based on the topology, we have that the degree distribution of nodes in the TV show social network follows a power-law distribution $f(x) = -52 x^{-0.5} + 58$ ($R^{2} = 0.98$), indicating that the TV show social network can be regarded as a scale-free network\footnote{http://en.wikipedia.org/wiki/Scale\_free\_network}.


\subsection{Social Networks of Viewers and Actors}

An online microblogging service also allows us to explore the social relationship between the microblog users who have posted microblogs on the TV shows.

\subsubsection{Social Topology Among Viewers.}

\begin{figure}[h]
    \begin{minipage}[t]{.48\linewidth}
        \centering
            \includegraphics[width = \linewidth]{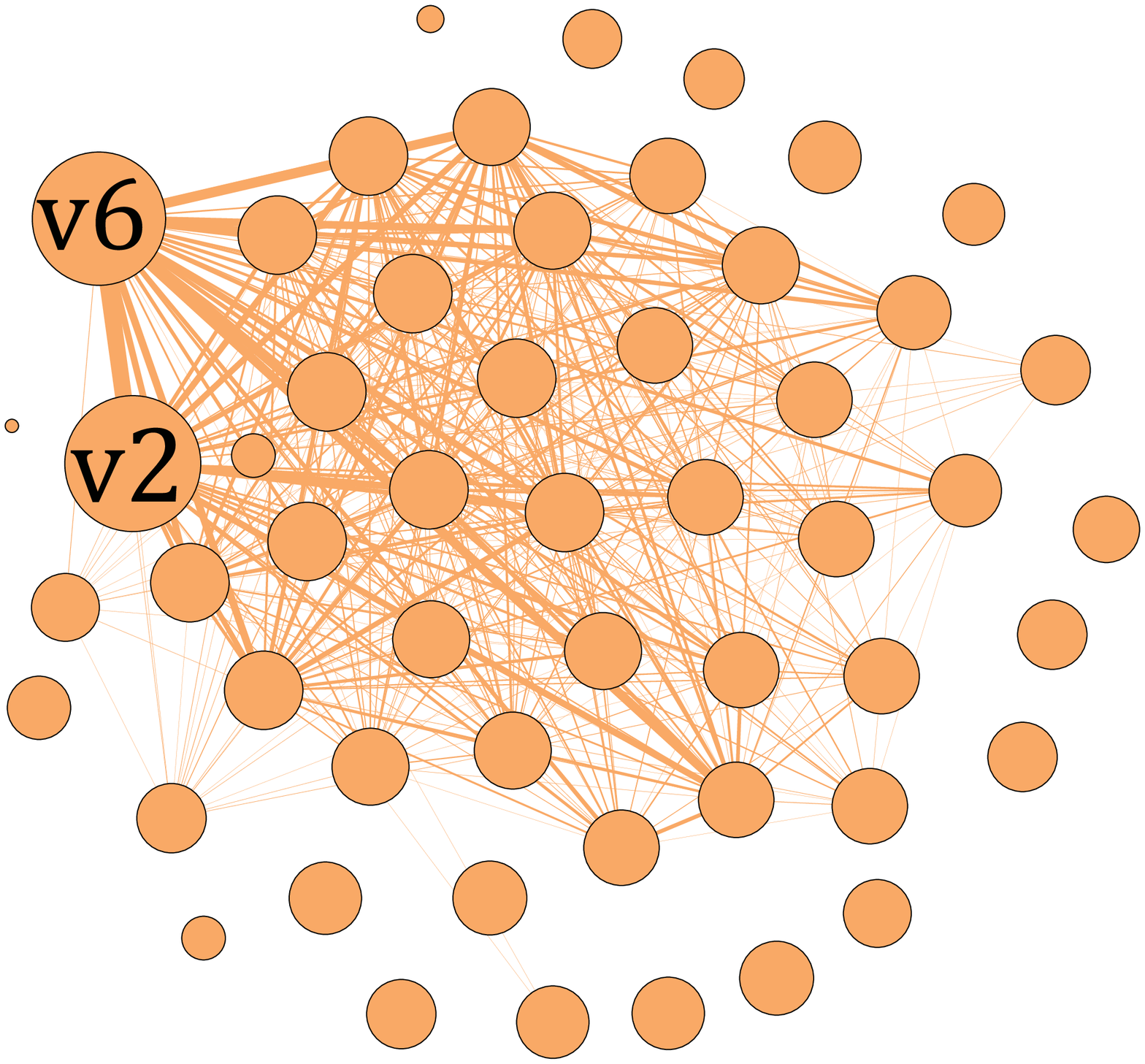}
        \caption{Communities in the social network of TV shows.}
        \label{fig:community-p}
    \end{minipage}
    \hfill
    \begin{minipage}[t]{.48\linewidth}
        \centering
            \includegraphics[width = \linewidth]{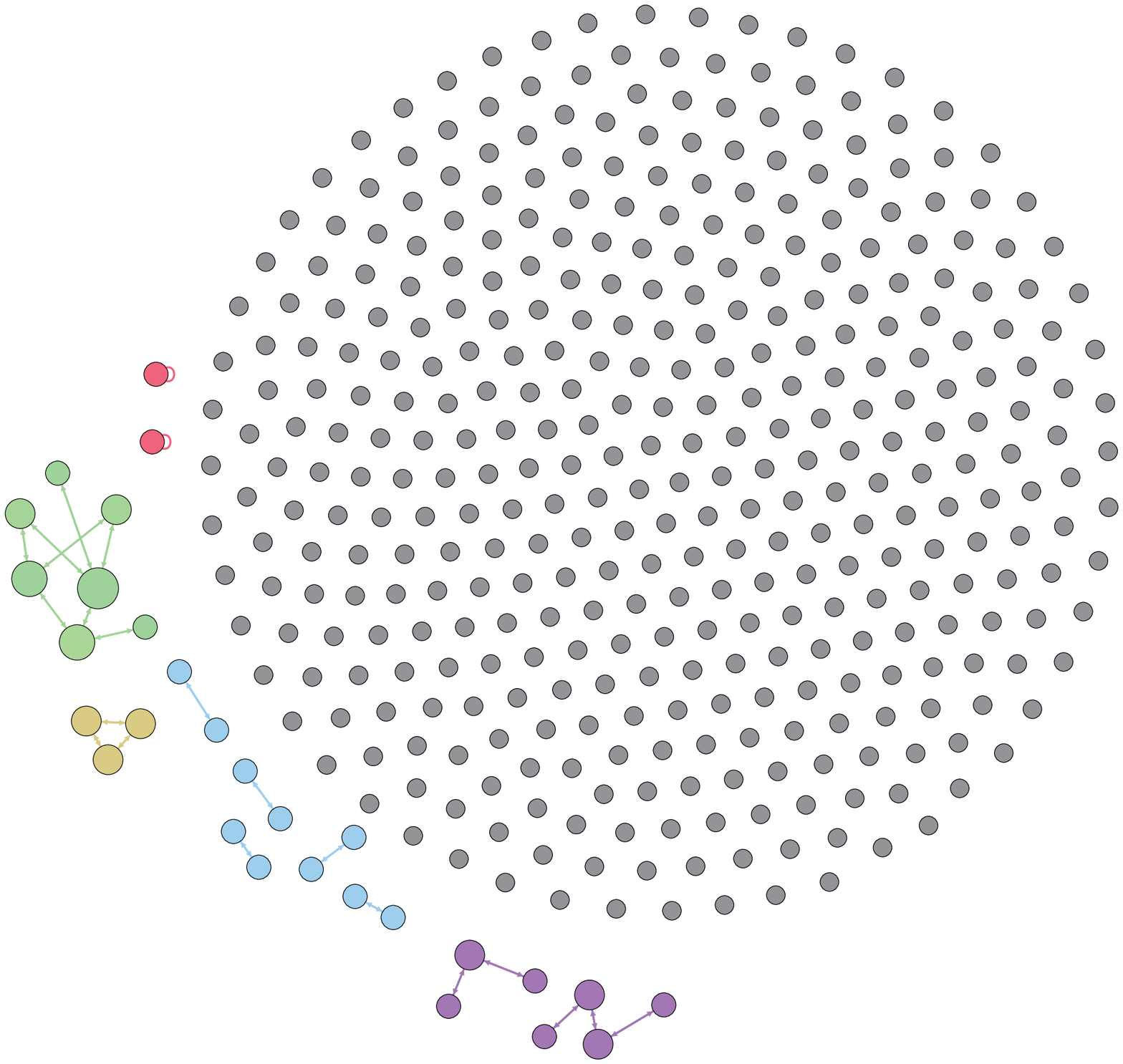}
        \caption{Social graph among users of TV show v1.}
        \label{fig:activityTopo}
    \end{minipage}
\end{figure}

We first study the following relationship among microblog users. As shown in Fig.~\ref{fig:activityTopo}, most of the users (about $76\%$) are isolated (with no social connection to other users), and a small fraction (about $24\%$) of them have social connections.


\subsubsection{Social Networks among Actors.}

\begin{figure}[t]
    \begin{minipage}[t]{.32\linewidth}
        \centering
            \includegraphics[width = 0.8\linewidth]{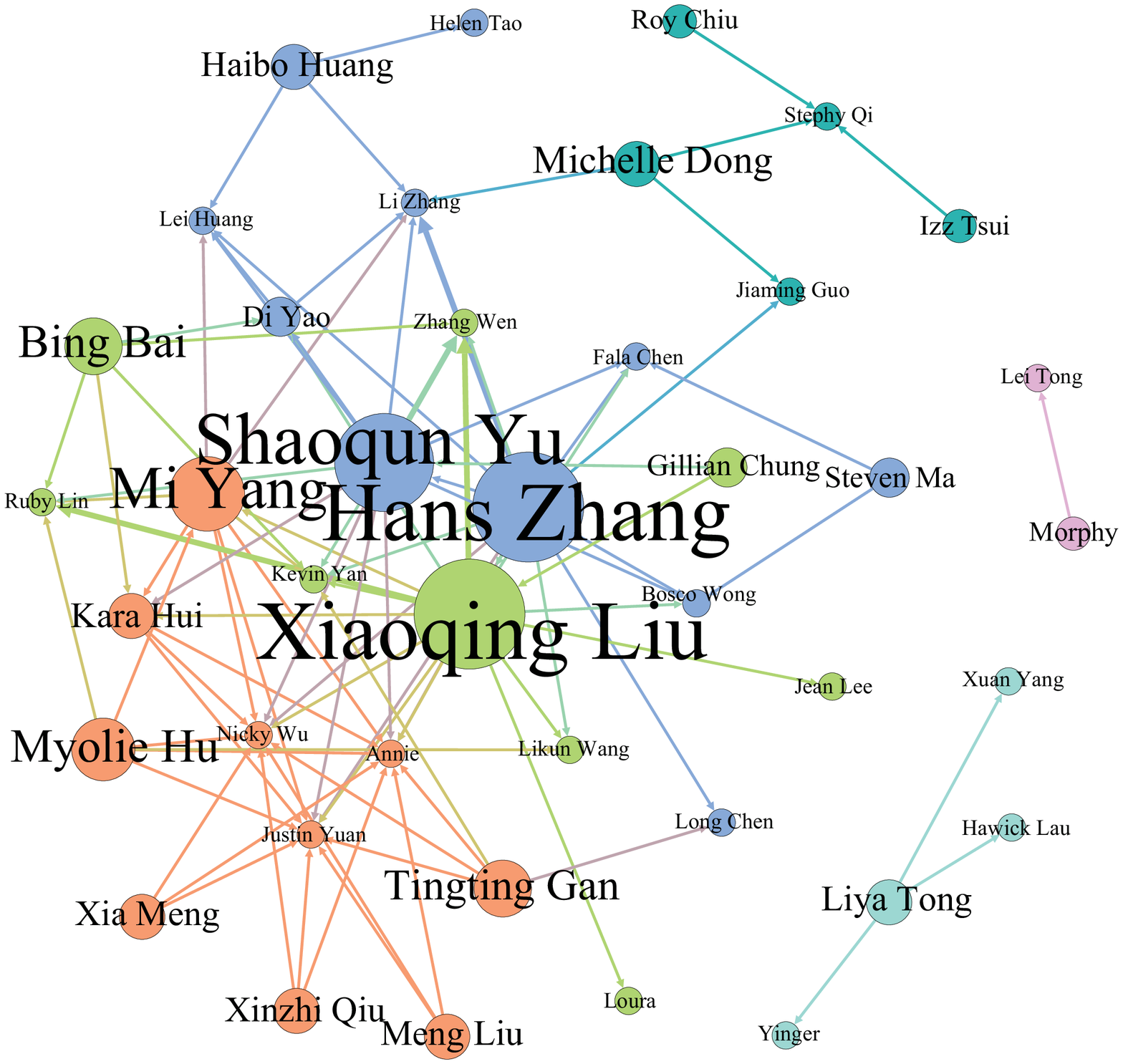}
        \centerline{\parbox[b]{\linewidth}{\scriptsize (a) A social network of actors posting to same shows.}}
    \end{minipage}
    \hfill
    \begin{minipage}[t]{.32\linewidth}
        \centering
            \includegraphics[width = 0.8\linewidth]{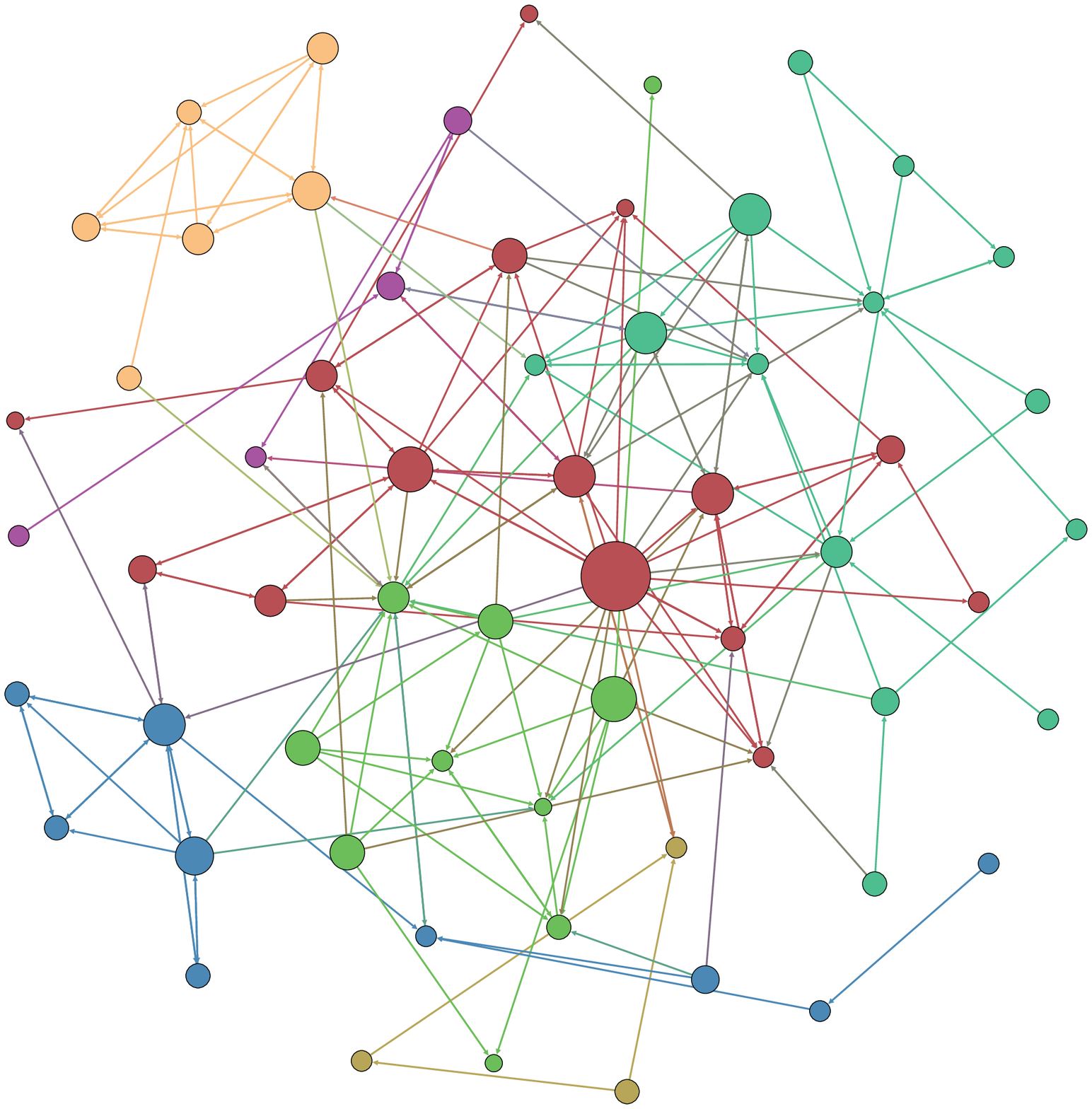}
        \centerline{\parbox[b]{\linewidth}{\scriptsize (b) A social network of actors with following relationship.}}
    \end{minipage}
    \hfill
        \begin{minipage}[t]{.32\linewidth}
        \centering
            \includegraphics[width = 0.8\linewidth]{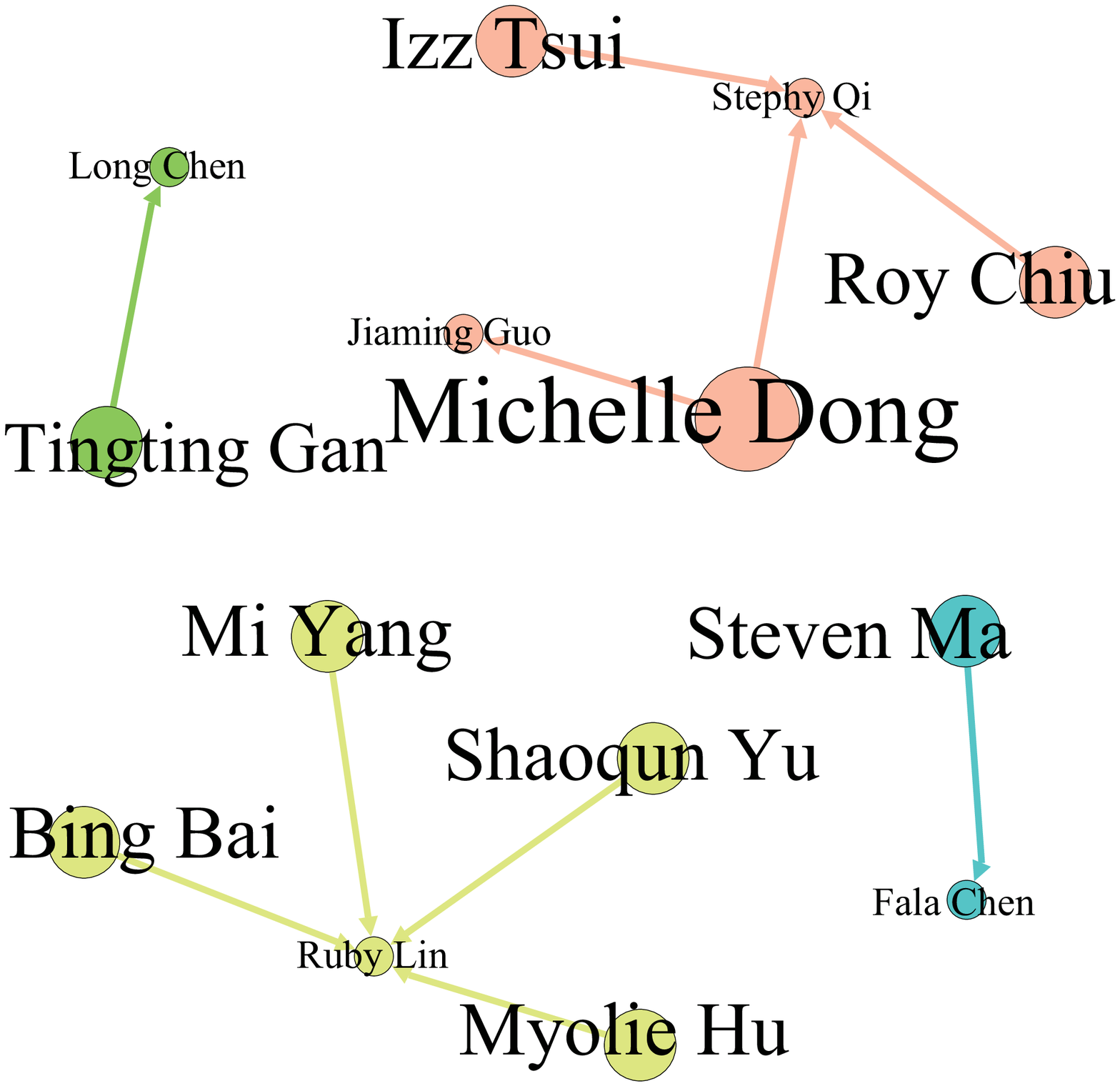}
        \centerline{\parbox[b]{\linewidth}{\scriptsize (c) A social network of intersection of (a) and (b)}}
    \end{minipage}
    \caption{Social networks among actors.}
    \label{fig:actorTopo}
\end{figure}

Today, online microblogging services are not only used by ordinary users, but also popularly used by celebrities and actors. We next study the social networks of actors of TV shows. Fig.~\ref{fig:actorTopo}(a) is a graph created based on microblogs posted by actors: there is an edge between two actors if one of them has posted a microblog on a TV show which involves the other. In this figure, we observe that some actors tend to help others to broadcast their shows, even they are not acting in those shows. In Fig.~\ref{fig:actorTopo}(b), we have plotted the following relationship between actors on the microblogging system. We observe that actors also share similar social topologies as ordinary users \cite{benevenuto2009characterizing}. In Fig.~\ref{fig:actorTopo}(c), we plot an intersection of (a) and (b), i.e., only edges that appear in both graph (a) and (b) are plotted. We observe very few edges remaining in this figure, indicating that it is very common that actors post microblogs on shows of other actors they are not following or friending on a microblogging system.


\subsubsection{Social Influence of Actors.}


Next, we study the social influence of actors to the shows they star in.

\begin{figure}[h]
    \begin{minipage}[t]{.48\linewidth}
        \centering
            \includegraphics[width = \linewidth]{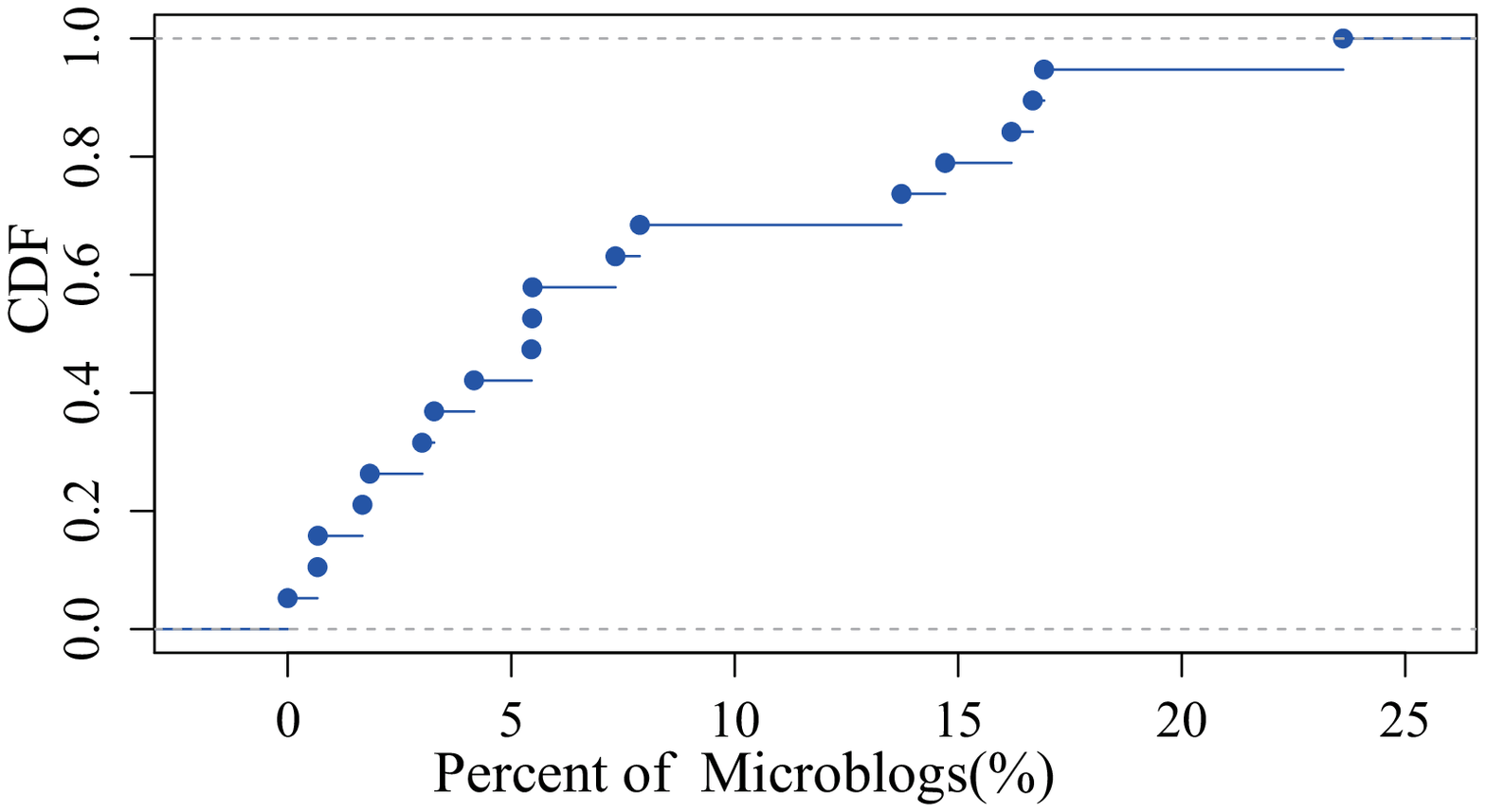}
        \centerline{\parbox[b]{\linewidth}{\scriptsize (a) CDF of percentage of microblogs posted by actors.}}
    \end{minipage}
    \hfill
    \begin{minipage}[t]{.48\linewidth}
        \centering
            \includegraphics[width = \linewidth]{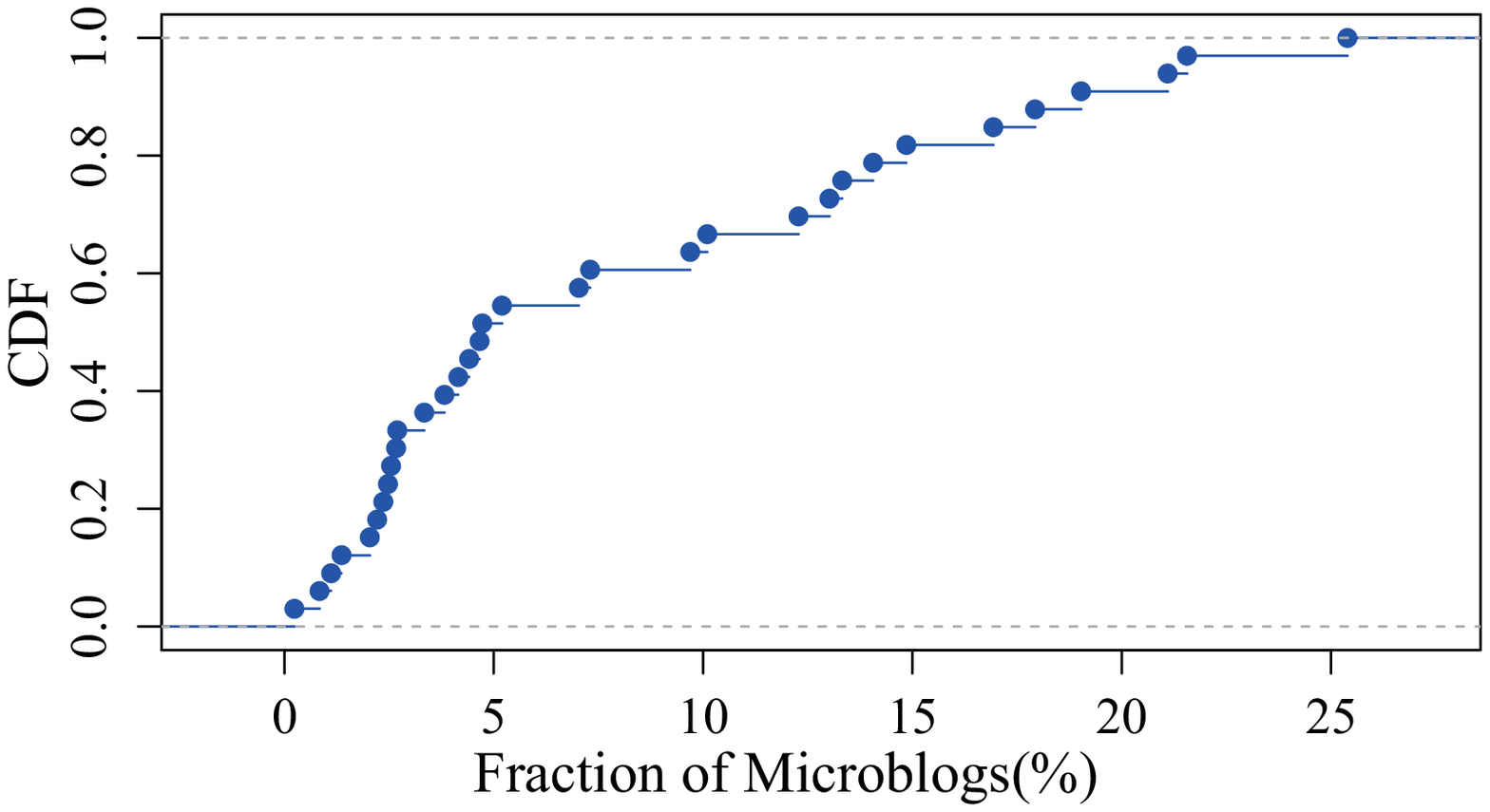}
        \centerline{\parbox[b]{\linewidth}{\scriptsize (b) CDF of fraction of microblogs posted by fans.}}
    \end{minipage}

    \caption{Social influence from actors of TV shows.}
    \label{fig:actorOwn}

\end{figure}

Fig.~\ref{fig:actorOwn}(a) shows the CDF of the fraction of microblogs by actors over all microblogs about a TV show. We have the following observations: (1) Actors post a large fraction of microblogs in a TV show; (2) Though more than half actors post more than $5\%$ TV-related microblogs, there is almost no actor posting more than $20\%$. Next, we study the microblogs posted by fans of the actors of a show. Fig.~\ref{fig:actorOwn}(b) plots the CDF of fractions of microblogs posted by fans of actors in a TV show. We observe that the fans of actors also input a large portion of microblogs for a TV show.

\begin{figure}[h]
    \centerline{\psfig{figure=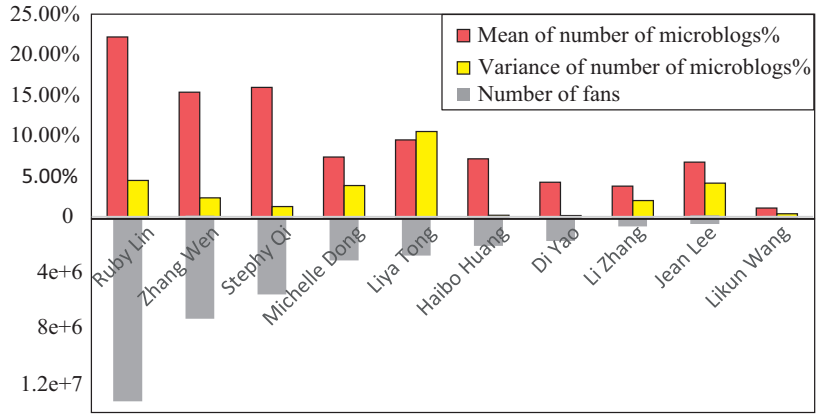,width = .9\columnwidth} }
    \caption{Actors' social influence and number of fans.}
    \label{fig:avgandvarinfluence}
\end{figure}

Finally, we study the heterogeneous influence of different actors. We calculate actors' average and variance influence on the shows they act in. The results are presented in Fig.~\ref{fig:avgandvarinfluence}, with their fans number. (1) We observe that for most of the actors, they tend to have a higher average influence if they have a larger number of fans. (2) The influence variance is generally small.

\subsection{Viewer Propagation across TV Shows}

Viewers of TV shows generally enjoy different shows at different periods. Based on our profiling scheme, we are able to study the propagation of users across different shows.

\subsubsection{Propagation of Viewers Over Time.}

\begin{figure}[h]
	\centerline{\psfig{figure=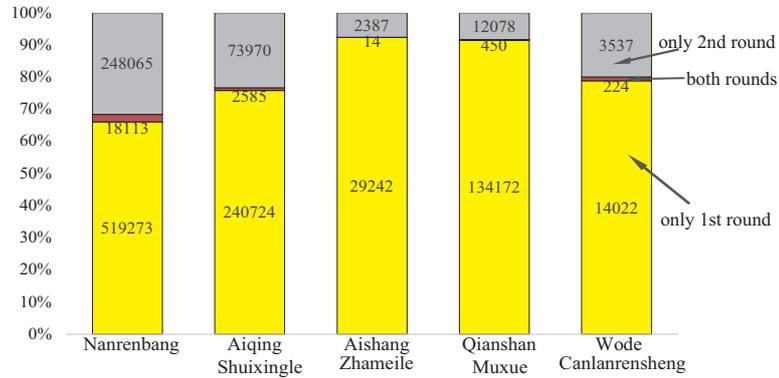,width = .85\columnwidth} }
	\caption{Propagation of viewers over time.}
    \label{fig:userOverlapDiffRounds}
\end{figure}

We study how users join the same show played at different times. In this experiment, shows are played in two different periods, denoted as the first round and second round. In Fig.~\ref{fig:userOverlapDiffRounds}, each bar plots the number of users in the first round, second round, and both rounds. We observe that though the number of users in both rounds accounts for a small proportion. This observation indicates that new users tend to join a TV show if the show is played again.

\subsubsection{Propagation of Viewers Across Shows.}

\begin{figure}[h]
    \begin{minipage}[t]{.48\linewidth}
        \centering
            \includegraphics[width = \linewidth]{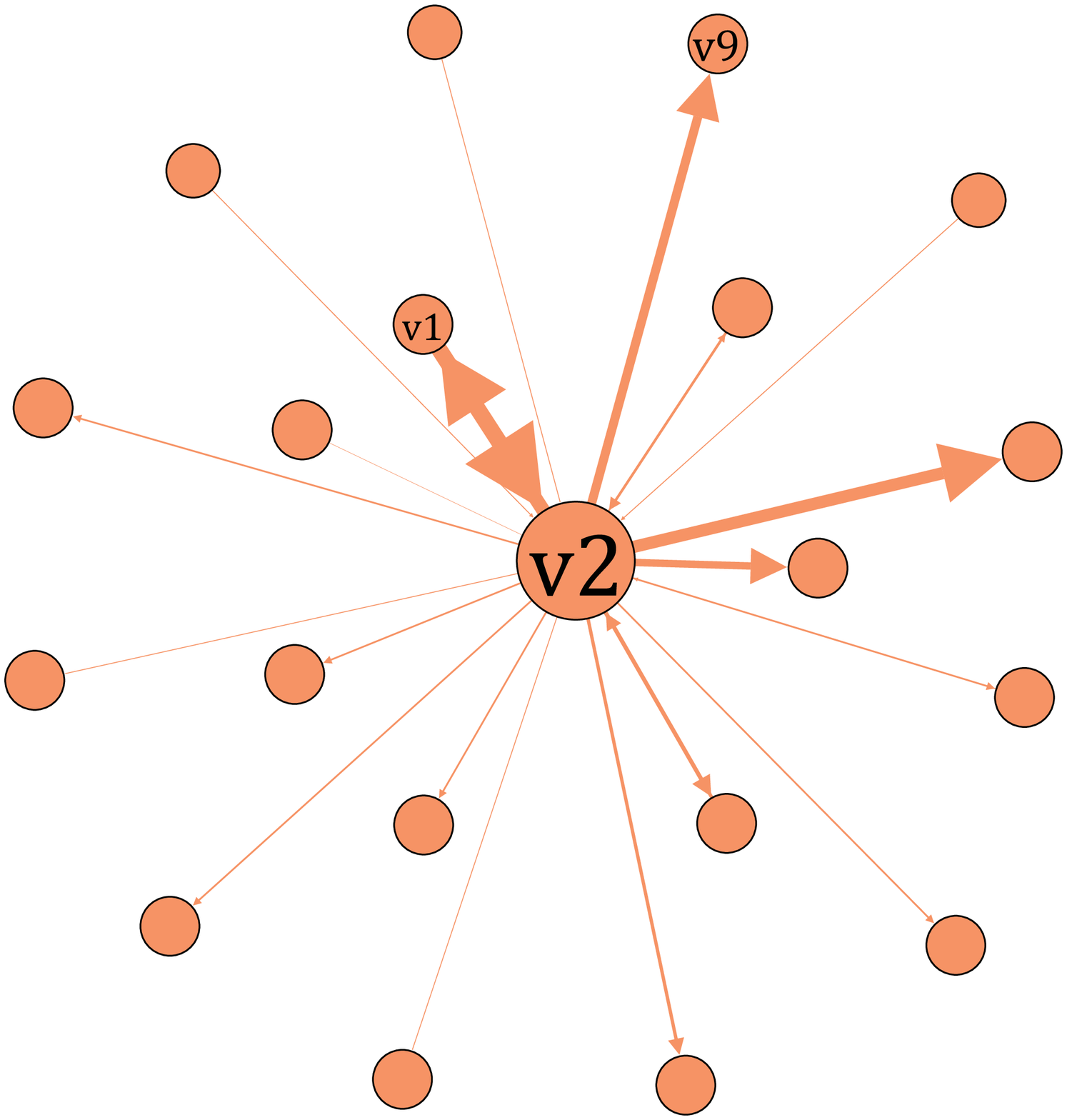}
        \centerline{\scriptsize (a)}
    \end{minipage}
    \hfill
    \begin{minipage}[t]{.48\linewidth}
        \centering
            \includegraphics[width = \linewidth]{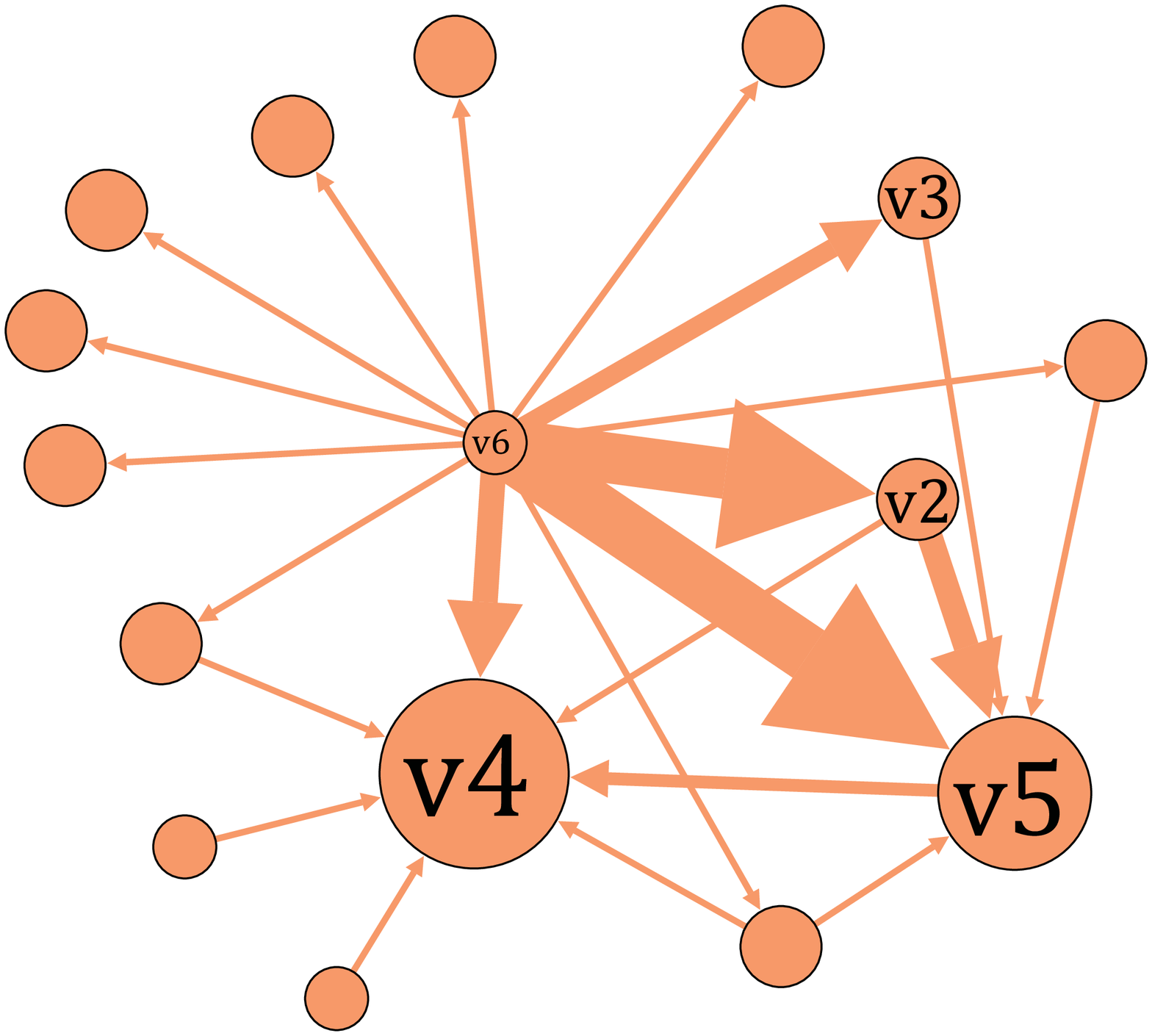}
        \centerline{\scriptsize (b)}
    \end{minipage}

    \caption{Propagation of microblogging users between TV shows.}
    \label{fig:userTran-3}
\end{figure}

Next, we study user propagation between TV shows. In Fig.~\ref{fig:userTran-3}, each node represents a TV show, and its diameter indicates the number of in-degrees. There is a directed edge between two nodes if users ``propagate'' from one TV show to another, i.e., microblogging users start to post microblogs to another from one TV show. The thickness of an edge indicates the number of users being propagating. In Fig.~\ref{fig:userTran-3}(a), we study a propagation case when a TV show v2 has a plummeting number of microblogging users. We observe that much more social-network users are propagating from v2 to other shows. Besides, we observe that among these propagation users, a major fraction of them are moving to TV show v1. The reason is that the content provider V began to promote v1 two days before the propagation, indicating that users can be highly influenced by new TV shows published.

In Fig.~\ref{fig:userTran-3}(b), we study another type of propagation. This figure illustrates a case when v6 is about to finish (delivering its last episode)£¬ and v4 and v5 are the newly released shows scheduled on the next day. We observe that a dominate fraction of users are propagating from v6 to v5. The reason is that v6 and v5 are in the same TV category. On the other hand, v4 is of a different TV category, and we observe that much fewer users are propagating from v6 to v4. This observation indicates that users tend to propagate between TV shows with similar content.

\section{Concluding Remarks}
\label{sec:conclusion}

In this paper, we study the profiling of TV shows using online microblogging services, in a sense to provide users with a socialized profile of TV shows. It is challenging to profile TV shows with microblogs since TV shows are generally offline, and an effective information retrieval scheme is in demand to find the most useful microblogs for the profiling; meanwhile, information on a microblogging system is noisy for the profiling and an efficient profiling scheme is in demand. To address these two challenges, we propose a joint actor and topic information retrieval scheme for searching microblogs to profile a particular TV show. Based on that, we propose \emph{MAP}, a microblog-assisted profiling framework, in which a TV show is profiled from the aspects of user, content, social relationship and social propagation. Using \emph{MAP}, we profile real-world TV shows and present several analysis results, which are not only interesting to the ordinary viewers, but also important to content providers for better operation.

\enlargethispage{60mm}

\subsubsection*{Acknowledgments.} This research is supported in part by NSFC under Grant No.~61272231, 61472204, and 61402247, SZSTI under Grant No.~JCYJ20140417115840259, and Beijing Key Laboratory of Networked Multimedia. We thank Tsinghua-Tencent Joint Lab and BesTV for providing the traces used our study.

\small


\end{document}